\documentclass{style}

\def\Journal#1#2#3#4{{#1} {\bf #2} (#3) #4}

\def\NPB{{\em Nucl. Phys.}   {\bf B}}
\def\PLB{{\em Phys. Lett.}   {\bf B}}

\def\PRD{{\em Phys. Rev.}    {\bf D}}
\def\ZPC{{\em Z. Phys.}      {\bf C}}
\def\EJC{{\em Eur. Phys. J.} {\bf C}}

\newcommand{\pom}{{I\!\!P}}

\def\bit{\begin{itemize}}
\def\eit{\end{itemize}}

\def\figsize{0.60}

\begin{document}

\title{Diffractive Jet Production in DIS \\
-- Testing QCD Factorisation
}

\author{F.-P. Schilling \ [H1 Collaboration]}

\address{DESY, Notkestr. 85, D-22603 Hamburg, Germany\\
E-mail: fpschill@mail.desy.de}

\maketitle

\abstracts{ 
  A measurement \cite{paper,phd} is presented of diffractive
  dijet production in deep-inelastic scattering (DIS) interactions at
  HERA of the type $ep\rightarrow eXY$, where the $X$ system is
  separated by a large rapidity gap from a low-mass leading baryonic
  system $Y$. The data are consistent with QCD factorisation in
  diffractive DIS and yield tight constraints on both shape and
  normalisation of the diffractive gluon distribution.  The data are
  able to distinguish between various models for diffractive DIS and
  are in agreement with a 2-gluon exchange calculation at small
  $x_\pom$ values.  
}

\section{Overview}

At HERA, colour singlet exchange or {\em diffractive} processes are
studied in deep-inelastic $ep$ scattering (DIS), where the exchanged
photon with virtuality $Q^2$ provides a probe to determine the QCD
(i.e. quark and gluon) structure of diffractive exchange.  In
\cite{collins}, it was proven that QCD hard scattering factorisation
is valid in diffractive DIS, so that {\em diffractive parton
  distributions} $p_i^D$ in the proton can be defined as
quasi-universal objects.  The hypothesis of a factorising $x_\pom$
dependence ({\em Regge factorisation}) is often used in addition.

\begin{figure}
\centering
\begin{minipage}{0.45\linewidth}
\centering
  \epsfig{file=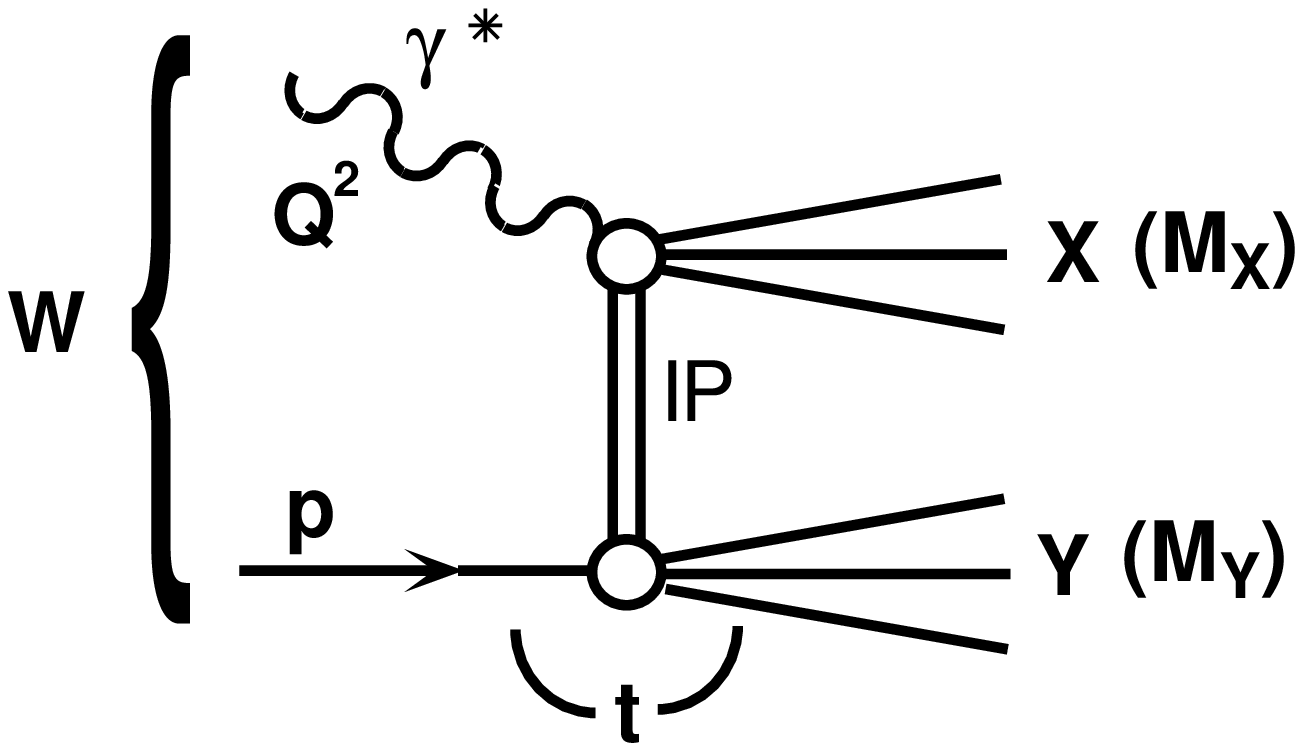,height=3.0cm,clip=}
\end{minipage}
\hfill
\begin{minipage}{0.45\linewidth}
\centering
  \includegraphics[clip,height=3.0cm]{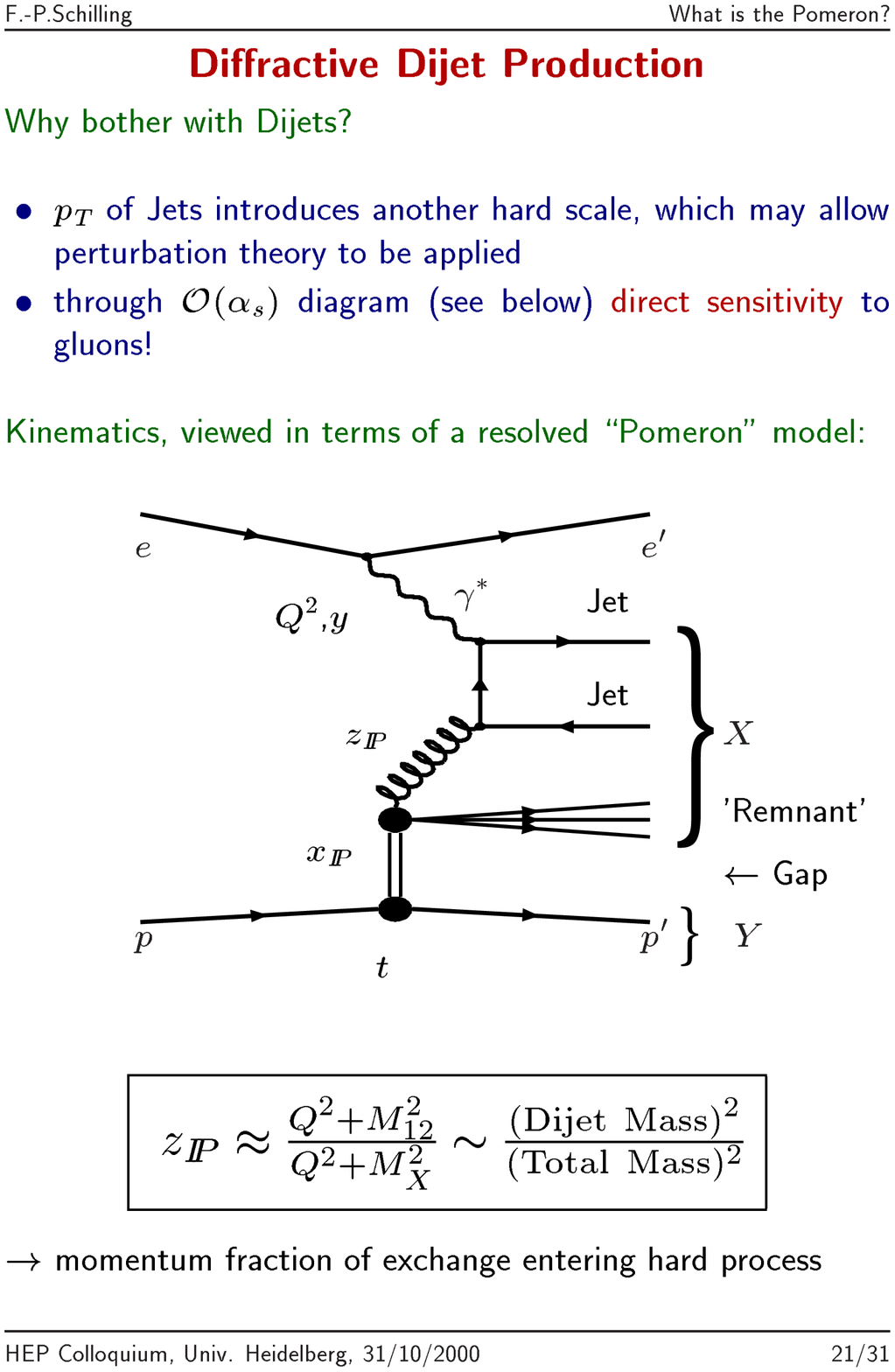} 
\end{minipage}
\caption{Inclusive diffractive scattering at HERA {\em (left)} and
diffractive dijet production {\em (right)}, viewed in a partonic picture.}
\label{fig2}
\end{figure}

Measurements of inclusive diffractive DIS in terms of the {\em
  diffractive structure function} $F_2^{D(3)}(x_\pom,\beta,Q^2)$
mainly constrain the diffractive quark distribution. By contrast,
diffractive dijet production is directly sensitive to the gluon
distribution $g^D(z,\mu^2)$ (Fig.~\ref{fig2}), which can be inferred
only indirectly from the scaling violations of $F_2^{D(3)}$. 
QCD factorisation can be tested by predicting the 
dijet cross sections using  the pdf's extracted from $F_2^{D(3)}$.

Furthermore, the predictions of a variety of phenomenological models
for diffractive DIS such as soft colour
neutralisation or 2-gluon exchange can be confronted with the dijet
cross sections.

\section{Data Selection and Cross Section Measurement}

The data sample corresponds to an integrated luminosity of
$\mathcal{L}=18.0 \rm\ pb^{-1}$ and was obtained with the H1 detector at
HERA. Dijet events were identified using the CDF cone algorithm and
diffractive events were selected by the requirement of a large
rapidity gap in the outgoing proton direction.  The kinematic range of
the measurement is $4<Q^2<80 \ \mathrm{GeV^2}$, $p^*_{T, jet}>4 \ 
\mathrm{GeV}$, $x_\pom<0.05$, $M_Y<1.6 \ \mathrm{GeV}$ and $|t|<1.0 \ 
\mathrm{GeV^2}$.  The cross sections were corrected for detector and
QED radiative effects and the systematic uncertainties, which dominate
the total errors, were carefully evaluated.

\section{Diffractive Parton Distributions}

Parton distributions for the diffractive exchange\footnote{ The
  assumption of Regge factorisation was found to be compatible with
  the data.}  were extracted from  DGLAP QCD fits to
$F_2^{D(3)}(x_\pom,\beta,Q^2)$ in \cite{h1f2d94}. The parton
distributions were found to be dominated by gluons 
($80-90\%$ of the exchange momentum).

\begin{figure}
\centering
\epsfig{width=\figsize\linewidth,file=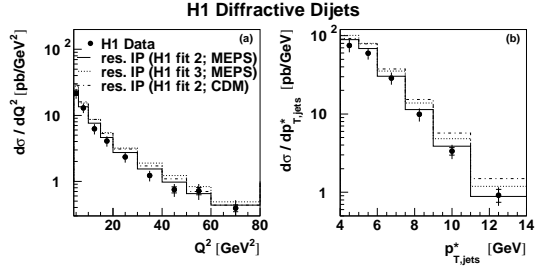}
\caption{Diffractive dijet cross sections as a 
function of {\em (a)}
$Q^2$ and {\em (b)} $p^*_{T,jets}$.}
\label{fig5ab}
\end{figure}

\begin{figure}
\centering
\epsfig{width=\figsize\linewidth,file=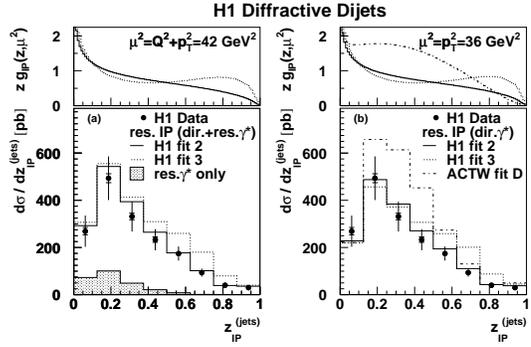}
\caption{Diffractive dijet cross sections as a function of 
$z_\pom^{(jets)}$. In {\em (a)} and {\em (b)}, the same data are
compared with different model predictions (see text).
}
\label{fig7}
\end{figure}

If these parton distributions, which evolve according to the DGLAP
equations, are used to predict the diffractive dijet cross sections, a
very good agreement is obtained (Fig.~\ref{fig5ab}). Fig.~\ref{fig7}
shows the measurement of the dijet cross section as a function of
$z_\pom^{(jets)}$, an estimator for the parton momentum fraction of
the diffractive exchange which enters the hard process (Fig.~\ref{fig2} 
right).  A
very good agreement in shape and normalisation is obtained if the {\em
  fit 2} parton distributions from \cite{h1f2d94} are used.
The {\em fit 3} parameterisation, in which
the gluon distribution is peaked towards high $z_\pom$ values, is 
disfavoured\footnote{The
corresponding gluon distributions are shown above the cross sections.}.
Using different factorisation scales ($\mu^2=Q^2+p_T^2$ in
Fig.~\ref{fig7}a, $\mu^2=p_T^2$ in Fig.~\ref{fig7}b) or including a
resolved virtual photon contribution (Fig.~\ref{fig7}a) 
in the model prediction does not
alter these conclusions.  The dijet data thus strongly support the validity of
QCD factorisation in diffractive DIS and give tight constraints on the
diffractive gluon distribution in both shape and normalisation.

\begin{figure}
\centering
\epsfig{width=\figsize\linewidth,file=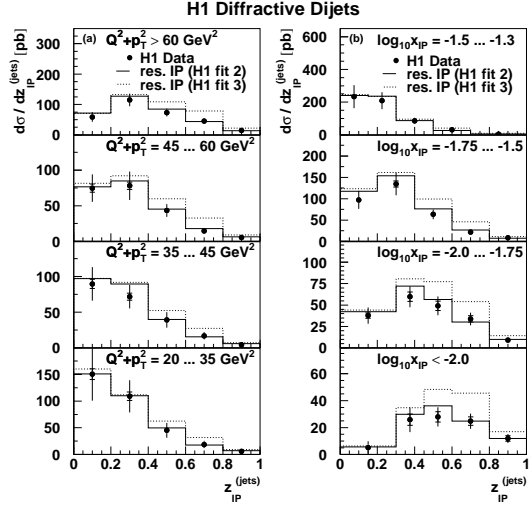}
\caption{Diffractive dijet cross sections as a function 
of $z_\pom^{(jets)}$ 
{\em (a)} in four bins of $\mu^2=Q^2+p_T^2$ and
{\em (b)} in four bins of $x_\pom$.}
\label{fig8}
\end{figure}

\begin{figure}
\centering
\epsfig{width=\figsize\linewidth,file=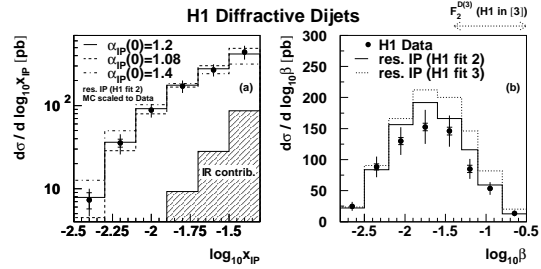}
\caption{Diffractive dijet cross sections as a function of 
{\em (a)} $\log_{10} x_\pom$ and {\em (b)} $\log_{10} \beta$.}
\label{fig6}
\end{figure}

The measured $z_\pom^{(jets)}$ cross sections in bins of the scale
$\mu^2=Q^2+p_T^2$ (Fig.~\ref{fig8}a) are in good agreement with the
prediction based on a DGLAP evolution of the diffractive parton
distributions.
The $z_\pom^{(jets)}$ cross sections in bins of
$x_\pom$ (Fig.~\ref{fig8}b) demonstrate consistency with Regge
factorisation. 

In a Regge framework, the energy dependence of the cross section is
determined in terms of an effective {\em pomeron intercept}
$\alpha_\pom(0)=1.17\pm0.07$ (stat.+syst.) from the $x_\pom$ cross
section (Fig.~\ref{fig6}a), consistent with the result from
\cite{h1f2d94}.  The cross section as a function of $\beta$ is shown
in Fig.~\ref{fig6}b.

\section{Soft Colour Neutralisation Models}

\begin{figure}
\centering
\epsfig{width=\figsize\linewidth,file=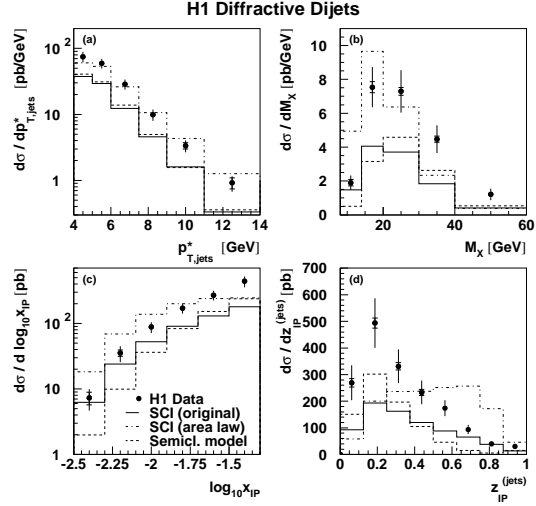}
\caption{Diffractive dijet cross sections, compared with 
the predictions of soft colour neutralisation models (see text).
}
\label{fig10}
\end{figure}

In Fig.~\ref{fig10}, the cross sections are compared with
models based on the ideas of soft colour neutralisation to
produce large rapidity gaps.  These are the original
version of the `soft colour interactions' model (SCI) \cite{sci},
the improved version of SCI based on a generalised area law
\cite{scinew} and the `semi-classical model' \cite{semicl}.
The original SCI and the semi-classical models give good descriptions
of the differential distributions.
However, none of  these models is yet able to simultaneously reproduce
shapes and normalisations of the dijet cross sections.


\section{Colour Dipole and 2-gluon Exchange Models}

\begin{figure}
\centering
\epsfig{width=\figsize\linewidth,file=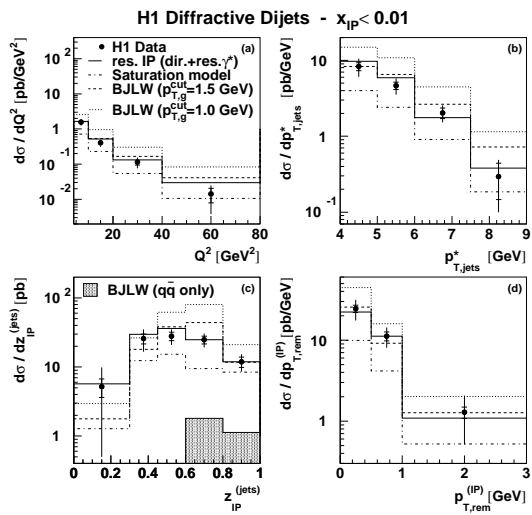}
\caption{Diffractive dijet cross sections for
$x_\pom<0.01$, compared with the predictions of 2-gluon 
exchange models (see text).
}
\label{fig11}
\end{figure}

Models for diffractive DIS based on the diffractive scattering of
$q\bar{q}$ or $q\bar{q}g$ photon fluctuations off the proton by
2-gluon exchange are confronted with the data in Fig.~\ref{fig11} for
the limited kinematic range of $x_\pom<0.01$, where contributions from
quark exchange can be neglected. The `saturation model' \cite{sat},
which takes only $k_T$-ordered configurations of the final
state partons into account,
reproduces the shapes of the differential distributions, but
underestimates the cross sections by a factor of two.  The model of
Bartels et al. \cite{bartels}, in which also non-$k_T$-ordered
configurations are taken into account,
is found to be in reasonable agreement
with the data if a free parameter $p_{T,g}^{cut}$ is fixed to $1.5
\rm\ GeV$\footnote{$p_{T,g}^{cut}$ corresponds to the minimum $p_T$
  of the final state gluon in the case of $q\bar{q}g$ production.}.

\section{Conclusions}

Diffractive dijet production has been shown to be a powerful tool
to gain insight into the underlying QCD dynamics of diffraction, in
particular the role of gluons. Factorisable, gluon-dominated diffractive
parton distributions successfully describe diffractive jet production 
and inclusive diffraction in DIS at the same time, in agreement with QCD
factorisation.


\end{document}